\documentstyle[11pt]{article}
\input{epsf}

\font\title=cmr17
\font\author=cmr12
\font\thanks=cmr7

\newtheorem{theorem}{Theorem}

\newtheorem{lemma}{Lemma}

\begin{document}

\begin{titlepage}

\begin{center}
{\Large Universal Lossless Data Compression}
\end{center}
\begin{center}
{\Large Via Binary Decision Diagrams}
 \end{center}
\bigskip

\centerline{\author J.\ Kieffer, P.\ Flajolet, and E-h.\ Yang}

\vskip 5.0truein
\vrule height .01 in width 2.5 in \par
{\thanks This work was supported in part by  National Science Foundation
Grants NCR-9508282 and NCR-9902081, and by the Natural Sciences and
Engineering Research Council of Canada, and was presented at the
IEEE International Symposium on Information Theory, Sorrento, Italy,
June 25--30, 2000. }  \par

{\thanks John Kieffer is with the Department of Electrical \& Computer
Engineering, University of Minnesota, Room 4-174 Keller Hall, 200 Union
Street SE, Minneapolis, MN 55455, USA. E-mail: kieffer@umn.edu} \par

{\thanks Philippe Flajolet is with INRIA-Rocquencourt, B. P.\ 105,
78153 Le Chesnay Cedex, France.}

{\thanks En-hui Yang is with the Department of Electrical \& Computer
Engineering, University
of Waterloo, Waterloo, Ontario, CA N2L 3G1. E-mail: ehyang@bbcr.uwaterloo.ca}
\par

\end{titlepage}

\centerline{\bf Abstract}
A binary string of length $2^k$ induces the Boolean function of $k$ variables
whose Shannon expansion is  the given binary string. This Boolean function
then is representable via a unique reduced ordered binary decision diagram (ROBDD).
The given binary string is fully recoverable from this ROBDD. We exhibit a lossless
data compression algorithm in which a binary string of length a power of two
is compressed via compression of the ROBDD associated to it as described above.
 We show that when  binary
strings of length $n$ a power of two are compressed via this algorithm,  the
maximal pointwise redundancy/sample with respect to any $s$-state binary
information source has the upper bound $(4\log_2s+16+o(1))/\log_2n $. To establish this result,
we exploit a result of Liaw and Lin stating that the ROBDD representation
of a Boolean function of $k$ variables contains a number of vertices
on the order of $(2+o(1))2^{k}/k$.

\vspace{3ex}


\newpage

\section{Introduction}
\label{sec1}
\setcounter{equation}{0}

Let $S({\rm dyadic})$ denote the set of all binary strings $x$ such that
\begin{itemize}
\item The length of $x$ is a power of two.
\item The substring of $x$ which forms the left half of $x$ does not
coincide with the substring of $x$ which forms the right half of $x$.
\item $x$ contains at least one entry of $1$ and at least one entry
of $0$.
\end{itemize}
Let $x \in S({\rm dyadic})$ and let $k$ be the logarithm to the base two of the
length of $x$. In a natural way, $x$ induces a Boolean function $f_x$
of $k$ variables. The function $f_x$ maps the set $\{0,1\}^k$ into the
set $\{0,1\}$, and can be defined as follows. Let $u_1, u_2, \dots, u_{2^k}$
be the lexicographical ordering of all binary strings of length $k$.
For each $i=1,2,\dots, 2^k$, define $f_x(u_i)$ to be the $i$-th
coordinate of $x$. Following \cite{bryant1} \cite{bryant2}, the
Boolean function $f_x$ can be represented by a
directed acyclic graph called a {\it reduced ordered binary decision
diagram} (ROBDD). Since the Boolean function $f_x$ can be recovered from
its ROBDD representation, $x$ can also be recovered from this representation. This means that
we  can losslessly compress a string $x \in S({\rm dyadic})$ by compressing
the ROBDD representation of the Boolean function $f_x$ induced by $x$. It
is the purpose of this note to investigate the compression performance
that is achievable by such a compression algorithm (we obtain a redundancy bound).\par

There are public domain software packages (some of them on the Internet) for
computing the ROBDD representation of a Boolean function. Such a package would
be easily adaptable in order to provide compression of a data string in $S({\rm dyadic})$
according to the ROBDD-based compression algorithm that we shall present.

\section{ROBDD's Representing Data Strings}
Define $\cal G$
to be the set of all finite graphs $G$ such that
\begin{description}
\item[(a)] $G$ is directed and acyclic.
\item[(b)] $G$ has a unique nonterminal vertex $V_G^r$ such that for every other
vertex
$V$ of $G$, there is at least one directed path leading from $V_G^r$ to $V$. The
vertex $V_G^r$
is called the root vertex of $G$.
\item[(c)] There are exactly two terminal vertices of $G$, which shall be denoted
$T_G^0$ and $T_G^1$, respectively.
\item[(d)] From each nonterminal vertex of $G$, there emanate exactly two edges,
one of which is labelled ``$0$'' and the other of which is labelled ``$1$''.
These edges terminate at different vertices (i.e., $G$ has no multiple edges).
\item[(e)] Each vertex $V$ of $G$ carries a positive integer label $L(V)$
which we shall call the level of $V$. The levels of the vertices satisfy
the properties:
\begin{itemize}
\item $L(V_G^r) = 1$.
\item $L(T_G^0) = L(T_G^1)$.
\item If $V_1, V_2, \dots, V_j$ are the vertices visited in order along any directed path
in $G$, then $L(V_1) < L(V_2) < \dots < L(V_j)$. (Note: The labels
$L(V_1), L(V_2), \dots, L(V_j)$ are not necessarily consecutive
integers.)
\end{itemize}
\end{description}
\par
{\it Example 1.} The graph given in Figure 1, in which each vertex is
labelled by its level, is seen to satisfy the properties
(a)-(e). Therefore, this graph is a member of $\cal G$.

\par
Let $G \in {\cal G}$. Let $V(G)$ be the set of vertices of
$G$. Let $\{0,1\}^+$ denote the set of all binary strings of finite positive length.
We define $\phi_G$ to be the unique mapping from $V(G)$ into $\{0,1\}^+$ such that
\begin{itemize}
\item $\phi_G(T_G^0) = 0$ and $\phi_G(T_G^1) = 1$.
\item If $V$ is a nonterminal vertex of $G$, if the edge labelled $0$ emanating from
$V$ terminates at vertex $V_0$, and the edge labelled $1$ emanating from $V$
terminates at vertex $V_1$, then
$$\phi_G(V) = \phi_G(V_0)^{(2^{L(V_0)-L(V)-1})}\phi_G(V_1)^{(2^{L(V_1)-L(V)-1})}$$
\end{itemize}
(Notation: If $y$ is a binary string and $j$ is a positive integer, then $y^j$
denotes the binary string obtained by concatenating together $j$ copies of $y$.
If $y_1$ and $y_2$ are binary strings, then $y_1y_2$ denotes the binary string
obtained by concatenating $y_2$ onto the right end of $y_1$.)
\par
{\it Example 2.} Let $A_1, A_2, \dots, A_{16}$  denote the sixteen vertices of the graph $G$ in Figure 1,
as indicated in Figure 2. (This is a ``canonical ordering'' of the vertices of $G$, which shall
be explained later.)
Since $A_8 = T_G^0$ and $A_{16} = T_G^1$, we have
\begin{eqnarray*}
\phi_G(A_8) & = & 0\\
\phi_G(A_{16}) & = & 1\\
\phi_G(A_9) & = & \phi_G(A_8)\phi_G(A_{16}) = 01\\
\phi_G(A_{10}) & = & \phi_G(A_8)^2\phi_G(A_{16})^2 = 0011\\
\phi_G(A_{11}) & = & \phi_G(A_9)\phi_G(A_{16})^2 = 0111\\
\phi_G(A_{12}) & = & \phi_G(A_8)^4\phi_G(A_{16})^4 = 00001111\\
\phi_G(A_{13}) & = & \phi_G(A_9)^2\phi_G(A_{16})^4 = 01011111\\
\phi_G(A_{14}) & = & \phi_G(A_{10})\phi_G(A_{16})^4 = 00111111\\
\phi_G(A_{15}) & = & \phi_G(A_{11})\phi_G(A_{16})^4 = 01111111\\
\phi_G(A_4) & = & \phi_G(A_8)^8\phi_G(A_9)^4 = 0000000001010101\\
\phi_G(A_5) & = & \phi_G(A_{10})^2\phi_G(A_{11})^2 = 0011001101110111\\
\phi_G(A_6) & = & \phi_G(A_{12})\phi_G(A_{13}) = 0000111101011111 \\
\phi_G(A_7) & = & \phi_G(A_{14})\phi_G(A_{15}) = 0011111101111111\\
\phi_G(A_2) & = & \phi_G(A_4)\phi_G(A_5) = {\rm length} \; 32 \; {\rm string}\\
\phi_G(A_3) & = & \phi_G(A_6)\phi_G(A_7) = {\rm length} \;  32\; {\rm string}\\
\phi_G(A_1) & = & \phi_G(A_2)\phi_G(A_3) = {\rm length }\; 64\; {\rm string}
\end{eqnarray*}
\par
The following is clear from the definition of $\phi_G$ and Example 2.\par
\begin{lemma} Let $G$ be any graph in $\cal G$. Suppose $L(T_G^0) = L(T_G^1) = k+1$.
Then, for each vertex $V$ of $G$, the length of $\phi_G(V)$ is $2^{k+1-L(V)}$.
In particular, the length of $\phi_G(V_G^r)$ is $2^k$.
\end{lemma}
\par
{\it Definition.} We define ${\cal G}^*$ to be the set of all graphs $G \in {\cal
 G}$
such that the mapping $\phi_G$ is one-to-one. \par

\begin{lemma} The following statements hold:
\begin{description}
\item[(a)] For any $G \in {\cal G}^*$, the binary string $\phi_G(V_G^r)$ is a member
of $S({\rm dyadic})$.
\item[(b)] For each $x \in S({\rm dyadic})$, there is a unique
$G \in {\cal G}^*$ such that $\phi_G(V_G^r) = x$. In the language of \cite{bryant1}
\cite{bryant2}, this unique graph $G$ is the unique ROBDD representing the Boolean function $f_x$.
\end{description}
\end{lemma}
\par
{\it Proof.} Part (a) is clear from Example 2. Part (b) (including the uniqueness of the
ROBDD representation) may be seen to be true by consulting the papers \cite{bryant1}
\cite{bryant2}. \par

{\it Notation.} For each $x \in S({\rm dyadic})$, we let $G_x$ denote the unique
graph in ${\cal G}^*$ which represents $x$ in the sense of Lemma 2(b). \par

{\it Example 3.} The graph $G$ in Figure 1 is $G_x$, where
$x \in S({\rm dyadic})$ is found from Example 2 by the calculation
\begin{eqnarray*}
x &   = & \phi_G(A_4)\phi_G(A_5)\phi_G(A_6)\phi_G(A_7)\\
& = & 0000000001010101\;0011001101110111\;0000111101011111\; 0011111101111111
\end{eqnarray*}\par

\section{Encoding Method}
For each $G \in {\cal G}^*$, we shall define in this section a binary codeword $\sigma(G)$
from which $G$ can be recovered.
Given $x \in S({\rm dyadic})$, we can then losslessly encode $x$ into the binary
codeword $\sigma(G_x)$.\par

We need the following notation. If $G$ is a graph in ${\cal G}^*$, and $V$ is a nonterminal vertex
of $G$, then the notation
$$V \to V_0, V_1$$
means that $V_0$ is the vertex of $G$ to which edge $0$ from $V$ leads, and $V_1$
is the vertex of $G$ to which edge $1$ from $V$ leads. \par

Fix $G \in {\cal G}^*$,
and let $j$ be the number of vertices of $G$. We define a canonical ordering of the
vertices of $G$. Let
$A_1, A_2, \dots, A_j$ be the enumeration of the vertices of $G$ which is uniquely
determined by the two properties
\begin{description}
\item[Property(i):] $A_1 = V^r$
\item[Property(ii):]
If $q_1 < q_2  < \dots < q_{j-2}$ are the integers in $\{1,2,\dots,j\}$
such that
$A_{q_1}, A_{q_2}, \dots, A_{q_{j-2}}$ are the nonterminal vertices of $G$, and if we write
\begin{eqnarray*}
A_{q_1} & \to & A_{r_1}, A_{s_1}\\
A_{q_2} & \to & A_{r_2}, A_{s_2}\\
& \dots & \\
A_{q_{j-2}} & \to & A_{r_{j-2}}, A_{s_{j-2}}
\end{eqnarray*}
then,
if we list the distinct entries of the sequence
$$(A_{r_1}, A_{s_1}, A_{r_2}, A_{s_2}, \dots, A_{r_{j-2}}, A_{s_{j-2}})$$
in order of their first left-to-right appearances in this sequence, we get
the list $A_2, A_3, \dots, A_j$.
\end{description}
\par
{\it Example 4.} The canonical ordering of the vertices of the graph $G$ in Figure 1
is given in Figure 2. We can determine this ordering by generating the following
relations one by one:
\begin{eqnarray}
A_1 & \to & A_2, A_3\nonumber\\
A_2 & \to & A_4, A_5\nonumber\\
A_3 & \to & A_6, A_7\nonumber\\
A_4 & \to & A_8,A_9\nonumber\\
A_5 & \to & A_{10},A_{11}\nonumber\\
A_6 & \to & A_{12},A_{13}\nonumber\\
A_7 & \to  & A_{14}, A_{15}\nonumber\\
A_9 & \to & A_8, A_{16}\nonumber\\
A_{10} & \to & A_8,A_{16}\nonumber\\
A_{11} & \to & A_9, A_{16}\nonumber\\
A_{12} &   \to & A_8,A_{16}\nonumber\\
A_{13} & \to & A_9, A_{16}\nonumber\\
A_{14} & \to & A_{10}, A_{16}\nonumber\\
A_{15} & \to & A_{11}, A_{16}\label{eq1}
\end{eqnarray}
Notice that in (\ref{eq1}), vertices $A_8$ and $A_{16}$ are missing from the left
hand sides. This means that $A_8$ and $A_{16}$ are the terminal vertices of the graph
in Figure 2. One of these vertices is equal to $T_G^0$ and the other is equal to
$T_G^1$.
We cannot determine which is the case from (\ref{eq1}) alone. We would need an extra
bit of information to determine which of the two possibilities
$$
\begin{array}{ll}
A_8=T_G^0&A_{16}=T_G^1\\
A_8=T_G^1&A_{16}=T_G^0\end{array}$$
holds.
\par

Let $G \in {\cal G}^*$, let $k$ be the positive integer such that $L(T_G^0) = L(T_G^1) = k+1$,
and let $A_1, A_2, \dots, A_j$ be the canonical ordering of the vertices of $G$.
We will generate strings $S_1, S_2, \dots, S_{k+1}$ in which
\begin{itemize}
\item $S_1 = A_1$, and each entry of each $S_i$ is a member of the set of symbols
$$\{A_m^q : m=1,2,\dots,j, \; \; q=1,2,\dots\}$$
\item The strings $S_1, S_2, \dots,  S_{k+1}$, taken together, allow one to build the
graph $G$ (except for the determination of which of the two terminal vertices
equals $T_G^0$, and which equals $T_G^1$, which takes one more bit of information,
as discussed above).
\item Each $S_i$ ($ i \geq 2$) is generated recursively from $S_{i-1}$ and certain
side information, and the side information from each recursive step is what
is encoded to form the overall codeword $\sigma(G)$. From $\sigma(G)$, the decoder
can then recursively generate the $\{S_i\}$, from which $G$ is obtained.
\end{itemize}
\par
Fix $i$, where $2 \leq i \leq k+1$.
The following procedure describes how
$S_i$ is recursively generated from $S_{i-1}$:
\begin{description}
\item[Step(i):] Write down the string $U$ consisting of
 the first appearances (from left to  right) of each distinct
symbol appearing in $S_{i-1}$.
\item[Step(ii):] For each entry of $U$ of form $A_m^q$, where $q > 1$,
write below that entry the entry $A_m^{q-1}$.
\item[Step(iii):] For each entry of $U$  of  form $A_m$, write down
below that entry the two
entries
$A_{m_0}^{q_0}, A_{m_1}^{q_1}$, where $A_{m_0}$, $A_{m_1}$ are the respective vertices
to which edges $0$ and $1$ from $A_m$ lead, and  $q_0$ and $q_1$ are
the positive integers
\begin{eqnarray*}
q_0 & = & L(A_{m_0}) - L(A_{m})\\
q_1 & = & L(A_{m_1}) - L(A_{m})
\end{eqnarray*}
\item[Step(iv):] Concatenate together the sequence of entries written below the
entries of $U$ in Steps (ii) and (iii). The resulting sequence is $S_i$.
\end{description}
\par

{\it Example 5.} For the graph $G$ in Figure 2, the strings $S_1, S_2, \dots, S_7$ are
as follows:
\begin{eqnarray*}
S_1 & = & A_1\\
S_2 & = & (A_2,A_3)\\
S_3 & = & (A_4,A_5,A_6,A_7)\\
S_4 & = & (A_8^4,A_9^3,A_{10}^2,A_{11}^2,A_{12},A_{13},A_{14},A_{15})\\
S_5 & = & (A_8^3,A_9^2,A_{10},A_{11},A_8^3,A_{16}^3,A_9^2,A_{16}^3,A_{10},A_{16}^3,A_{11},A_{16}^3)\\
S_6 & = & (A_8^2,A_9,A_8^2,A_{16}^2,A_9,A_{16}^2,A_{16}^2)\\
S_7 & = & (A_8,A_8,A_{16},A_{16})\\
\end{eqnarray*}
\par

Let $G \in {\cal G}^*$, let $k$ be the positive integer such that $L(T_G^0) = k+1$, and let
$A_1, A_2, \dots, A_j$ be the canonical ordering of the vertices of $G$.
One easily determines from $S_1,S_2,\dots,S_{k+1}$ the level of each vertex
$A_1,A_2,\dots,A_j$. For each $A_i$, find the unique $S_m$ such that $A_i$ is an
entry
of $S_m$. Then, $L(A_i) = m$. To illustrate, from $S_1, S_2, \dots, S_7$ in Example 5,
we determine that
$$
\begin{array}{llll}
L(A_1)=1&L(A_2)=2&L(A_3)=2&L(A_4)=3\\
L(A_5)=3&L(A_6)=3&L(A_7)=3&L(A_8)=7\\
L(A_9)=6&L(A_{10})=5&L(A_{11})=5&L(A_{12})=4\\
L(A_{13})=4&L(A_{14})=4&L(A_{15})=4&L(A_{16})=7
\end{array}
$$
Referring to Figure 2, we see that this assignment is correct.
\par
One also easily determines from $S_1,\dots,S_{k+1}$ where each edge of $G$ begins and ends.
For each nonterminal vertex $A_i$, find the unique $m < k+1$ such that $A_i$ is an entry
of $S_m$, and then look below in $S_{m+1}$ to find the corresponding two consecutive entries
$A_{i_0}^{q_0}, A_{i_1}^{q_1}$---vertices $A_{i_0}$ and $A_{i_1}$ are then the respective vertices at
which edges $0$ and $1$ from $A_i$ terminate. To illustrate, from $S_1, S_2, \dots, S_7$
in Example 5, we get the edge description given in (\ref{eq1}), which we see
is correct by referring to Figure 2.\par

For a graph $G \in {\cal G}^*$ such that $L(T_G^0) = L(T_G^1) = k+1$, we suppose
that the strings
$S_1, S_2, \dots, S_{k+1}$ have been generated. We now describe how these strings are encoded for
transmission to the decoder. The decoder already knows that $S_1 = A_1$. In addition to this,
 the decoder needs to know:
\begin{description}
\item[(a)] How to obtain $S_i$ from $S_{i-1}$, for each $i = 2,\dots,k+1$. This information
is transmitted to the decoder using $M_i$ codebits. In the sequel, we shall explain what
these $M_i$ codebits consist of.
\item[(b)] For the two symbols $A_{j_2}$ and $A_{j_2}$ comprising the entries of $S_{k+1}$, the decoder needs
to know which of these symbols equals $T_G^0$. This information is transmitted to the decoder
using one codebit.
\end{description}
\par
From the above description, we see that a total of $(M_2 + \dots + M_{k+1}) + 1$ codebits
is transmitted to the decoder by the encoder.
We need to further explicate Step (a) above, so that it is understood what $M_i$ is. To do this,
we need a number of definitions.\par

{\it Definition 1.} If $u = (u_1, u_2, \dots, u_J)$ is any nonempty
sequence of finite length over any alphabet $A$,
we define
$$H(u) \buildrel \Delta \over = \sum_{j=1}^J -\log_2\frac{n(u_j)}{J},$$
where, for each $a \in A$, $n(a)$ is the number of $1 \leq j \leq J$ for which $u_j = a$.
If $u$ is an empty sequence, we define $H(u) = 0$. The quantity $H(u)$ is important
for the following reason: If the set $\{u_1,u_2,\dots,u_J\}$ is known, and if the frequencies with which the
symbols in this set appear in $u$ are known, the sequence $u$
can be losslessly encoded using $ \lceil H(u)\rceil$ codebits. This is because there
are no more than $2^{H(u)}$ sequences having the known symbol frequencies.  \par

{\it Definition 2.} If $u$ is a sequence of finite length, $|u|$ denotes the length of $u$.\par

{\it Definition 3.} Let $u $ be any nonempty sequence of finite length over any alphabet.
 We
define $\tilde{u}$ to be the (possibly empty) sequence obtained from $u$ by striking out
each term of $u$ which is making its first left-to-right appearance in $u$.
For example, if
\begin{equation}
u = (a, a, b, a, b, c, b, b, c, a),\label{eq59} \end{equation}
we strike out the first, third, and sixth terms, obtaining
$$\tilde{u} = (a,a,b,b,b,c,a)$$
It could be that $u$ is empty. In this case, we define $\tilde{u}$ to be the empty sequence.
\par

{\it Definition 4.} If $u$ is a sequence of finite length such that $H(\tilde{u}) > 0$, we define
$h(u) = |u| + H(\tilde{u})$. If $u$ is a sequence of finite length such that $H(\tilde{u}) = 0$,
we define $h(u) = 0$. Here is why the quantity $h(u)$ is important: If the frequencies
with which the symbols appearing in $u$ are known, and if the list of these symbols
in order of first left-to-right appearance in $u$ is known, then the sequence $u$ can be
losslessly encoded using $\lceil h(u)\rceil$ codebits. To see this, one can encode $\tilde{u}$ using
$\lceil H(\tilde{u}) \rceil$ codebits. Then, one can obtain $u$ from $\tilde{u}$ with an additional
$|u|$ codebits (these additional codebits tell the decoder the positions in $u$ where the
first left-to-right appearances of the symbols in $u$ occur). This gives us a total of
$|u| + \lceil H(\tilde{u})\rceil = \lceil h(u)\rceil$ codebits. (We have assumed $H(\tilde{u}) = 0$.
The reader can treat the case $H(\tilde{u}) = 0$ separately.) For example, if $u$ is the
sequence in (\ref{eq59}), the additional $|u|$ codebits are $(1,0,1,0,0,1,0,0,0,0)$, the ones
indicating first appearances of $a,b,c$ in $u$ in positions $1,3,6$, respectively. \par

{\it Definition 5.} For each $2 \leq i \leq k+1$, we let $\hat{S}_i$ be the subsequence of
$S_i$ that arises from substituting for the distinct entries of $S_{i-1}$ of form $A_m$. (Recall that
each such entry of $S_{i-1}$ generates two entries of $S_i$.)\par

{\it Definition 6.} An entry of $\hat{S_i}$ of form $A_m^q$, where $A_m^{q+1}$ appears in $S_{i-1}$,
shall be called a Type I entry of $\hat{S_i}$. We let $\pi_i^1$ denote the subsequence of
$\hat{S}_i$ consisting of all the Type I entries of $\hat{S}_i$.\par

{\it Definition 7.} An entry of $\hat{S_i}$ of form $A_m^q$, where the symbol $A_m$ does not
appear in $S_{i-1}$, shall be called a Type II entry of $\hat{S}_i$. We let $\pi_i^2$
denote the subsequence of $\hat{S}_i$ consisting of all the Type II entries of $\hat{S}_i$.
Suppose that there are $r$ distinct entries of $\pi_i^2$, and that $A_m$ is the vertex of highest
index $m$ that has appeared in the sequences $S_1,S_2,\dots,S_{i-1}$. Then, if we list the
distinct entries of $\pi_i^2$ in order of their first left-to-right appearances in $\pi_i^2$,
this list will take the form
\begin{equation}
A_{m+1}^{q_1}, A_{m+2}^{q_2}, \dots,  A_{m+r}^{q_r}
\label{eq20}
\end{equation}
\par

{\it Definition 8.} We let $Q_i$ be the nonnegative integer consisting of the sum of all
the powers $q$ as $A_m^q$ ranges through all of the distinct terms of $\pi_i^2$. (In other words, referring to (\ref{eq20}), $Q_i$ is equal to $q_1 + q_2 + \dots + q_r$.)\par

With the above definitions, we can now stipulate that
\begin{equation}
M_i = |S_i| + |\hat{S_i}| + Q_i + \lceil H(\pi_i^1)\rceil +  \lceil H(\tilde{\pi}_i^2)\rceil,
\label{eq10}
\end{equation}
Here is how the different terms in $M_i$
arise:
\begin{description}
\item[(a.1)] Encoder transmits to decoder $|S_i|$ codebits to let the decoder know the
frequency with which each distinct element of $S_i$ appears.
\item[(a.2)] Encoder transmits to decoder $|\hat{S_i}|$ codebits so that the decoder will
know which entries of $\hat{S_i}$ are of Type I and which entries are of Type II.
\item[(a.3)] Encoder transmits to decoder $Q_i$ codebits so that the decoder will know
the powers $q$ appearing in the Type II entries $A_m^q$ of $\hat{S}_i$.
\item[(a.4)] The encoder transmits to the decoder
$\lceil H(\pi_i^1)\rceil$
codebits, which tell the decoder what $\pi_i^1$ is.
\item[(a.5)] The encoder transmits to the decoder $\lceil h(\pi_i^2)\rceil $ codebits,
which tell the decoder what $\pi_i^2$ is.
\end{description}
\par

{\it Definition.} We let $\sigma(G)$ be the binary codeword of length $(M_2+\dots +M_{k+1})+1$
obtained by concatenating together the codebits from Steps (a.1)-(a.5), (b) above.\par

{\it Example 6.} We explain how the decoder can obtain $S_5$ from $S_4$ in
Example 5.
Initially, the decoder will know that $S_5$ takes the form
$$S_5 = (A_8^3,A_9^2,A_{10},A_{11},\hat{S}_5),$$
where the entries of $\hat{S}_5$ have to be filled in. The decoder knows that the length
of $S_5$ is $12$. The decoder looks at the first $12$ codebits that are currently in its codebit buffer, to determine the frequencies of the distinct entries of $S_5$.
In this case, these $12$ codebits are
$$0,1,0,1,0,1,0,1,0,0,0,1$$
which tell the decoder that $A_8^3$ appears twice in $S_5$, $A_9^2$ appears twice in
$S_5$, $A_{10}$ appears twice in $S_5$,  $A_{11}$ appears twice in $S_5$, and an element
of form $A_{16}^q$, with $q$ unknown, appears four times in $S_5$. The decoder now knows
that $\pi_5^1$ is of length one and consists of one appearance of each of the symbols
$A_8^3, A_9^2, A_{10}, A_{11}$, and that $\pi_5^2$ is of length four and consists of four appearances of the symbol $A_{16}^q$. The next $|\hat{S}_5| = 8$ codebits in the decoder's buffer tell the decoder which entries of $\hat{S}_5$ are of Type I
and which are of Type II. In this case, these codebits are
$$(0,1,0,1,0,1,0,1),$$
which tell the decoder that the entries of $\hat{S}_5$ alternate between Type I entries
and Type II entries, starting with a Type I entry. The decoder now needs to determine
the power $q$ in the symbol $A_{16}^q$. In this case, $q=3$, and the decoder will know
this because the codebits
$$(0,0,1)$$
will appear at the start of the decoder's codebit buffer at this point. The
next
$\lceil H(A_8^3,A_9^2,A_{10},A_{11})\rceil = 8$
codebits tell the decoder that
$$\pi_5^1 = (A_8^3,A_9^2,A_{10},A_{11})$$
The decoder already
knows that
$$\pi_5^2 = (A_{16}^3,A_{16}^3,A_{16}^3,A_{16}^3),$$
so that, putting $\pi_5^1$ and $\pi_5^2$ together, the decoder has determined that
$$\hat{S}_5 = (A_8^3,A_{16}^3,A_9^2,A_{16}^3,A_{10},A_{16}^3,A_{11},A_{16}^3)$$

\section{Performance Bound}

Let $G \in {\cal G}^*$, and let $L(T_G^0) = k+1$. The binary codeword $\sigma(G)$
results
by encoding the sequences $S_2, S_3, \dots, S_{k+1}$, plus the
transmission of an extra codebit to signal the decoder which of the two terminal
vertices of $G$ is equal to $T_G^0$.
In this section, we want to upper bound the codeword length $|\sigma(G)|$, in order to see how good the
encoder is. \par
From the previous section, it can be seen that
\begin{equation}
|\sigma(G)| \leq 4[|S_1| + |S_2| + \dots + |S_{k+1}|] + \sum_{i=2}^{k+1}[\lceil H(\pi_i^1)\rceil +
\lceil H(\tilde{\pi}_i^2)\rceil]\label{eq61}\end{equation}
The only tricky part in obtaining this bound is the observation that
$$\sum_{i=2}^{k+1} Q_i \leq |S_2| + |S_3| + \dots + |S_{k+1}|$$
To see this, notice that if a Type II symbol $A_m^q$ appears in a sequence $S_i$, and $q > 1$, then the
$q-1$ symbols
$A_m^{q-1}, A_m^{q-2}, \dots, A_m$ appear in subsequent sequences $S_{i+1}, S_{i+ 2}, \dots$.
Summing the powers $q$ for all such symbols $A_m^q$, one must obtain a quantity
$Q_2 + \dots + Q_{k+1}$ upper bounded by $|S_2| + \dots + |S_{k+1}|$. \par

Let $x$ be the binary string of length $2^k$ represented by $G$ (i.e., $\phi_G(V_G^r) = x$).
Fix $i$ satisfying $2 \leq i \leq k+1$. From left to right, partition $x$ into disjoint substrings
of length $2^{k-i+2}$, and let $u_1, u_2, \dots, u_M$ be the list of distinct substrings
in this partition, listed in order of first left-to-right appearance in the partition.
For each $u_m$ in this list, let $u_m(L)$ denote the prefix of $u_m$ of length $2^{k-i+1}$,
and let $u_m(R)$ denote the suffix of $u_m$ of length $2^{k-i+1}$. (In other words, when
we bisect the string $u_m$, we obtain $u_m(L)$ on the left, and $u_m(R)$ on the right.)
Replace each $u_m$ in the sequence $ (u_1, \dots, u_M)$ for which $u_m(L) \not = u_m(R)$ by the
pair of strings $u_m(L), u_m(R)$; otherwise, if $u_m(L) = u_m(R)$, replace $u_m$ by
$u_m(L)$. These replacements yield a new sequence $v_i$ whose entries are substrings of $x$
of length $2^{k-i+1}$.
The following properties can be proved (see \cite{kief}).
\begin{description}
\item[Property 1:] The sequence ${S}_i$ has the same length as the sequence $v_i$.
\item[Property 2:] Writing
\begin{eqnarray*}
{S}_i & = & (q_1, q_2, \dots, q_{M})\\
v_i & = & (r_1, r_2, \dots, r_M)
\end{eqnarray*}
the sets $\{q_1, q_2, \dots, q_M\}$
and $\{r_1, r_2, \dots, r_{M}\}$ are of the same size, and there is a one-to-one mapping
$\alpha_i$ from the first set onto the second set in which
$$v_i  = (\alpha_i(q_1), \alpha_i(q_2), \dots, \alpha_i(q_{M}))$$
\item[Property 3:]  There is a partition $\Pi$ of $x$, and disjoint subsequences
$s^2, s^3, \dots, s^{k+1}$ of $\Pi$ (some of which may be empty), such that
$$s^i = \tilde{v}_i, \; \; 2 \leq i \leq k+1$$
\end{description}

{\it Definitions.} We let $\Lambda$ denote the family of
all mappings   $\lambda : \{0,1\}^+ \to (0,1]$ such that
for every sequence $u \in \{0,1\}^+$, and every partition $(u_1,u_2,\dots,u_r)$ of $u$
into nonempty substrings of $u$,
\begin{equation}
\lambda(u) \leq \lambda(u_1)\lambda(u_2)\dots\lambda(u_r)\label{eq1000}
\end{equation}
If $\lambda \in \Lambda $, we define
$$|\lambda| = \sup_{n=1,2,\dots}\sum_{u \in \{0,1\}^n} \lambda(u)$$

\begin{lemma} Let $\lambda$ be a function  in $\Lambda$ for which
$|\lambda| < \infty$. Let $G \in {\cal G}^*$, let $x$ be the binary
string of length $2^k$ represented by $G$, and let $S_1, S_2, \dots, S_k$ be the strings
defined for $G$ according to Section II. Then,
\begin{equation}
\sum_{i=2}^{k+1}[H(\pi_i^1) + H(\tilde{\pi}_i^2)] \leq \left ( \sum_{i=2}^{k+1} |S_i|\right ) \log_2 |\lambda| - \log_2\lambda(x)
\label{eq30}
\end{equation}
\end{lemma}
\par

{\it Proof.}  The sequences $\pi_i^1$ and $\tilde{\pi}_i^2$ are disjoint subsequences of
$\tilde{S}_i$. Applying Property 2, we have
$$H(\pi_i^1) + H(\tilde{\pi}_i^2) \leq H(\tilde{S}_i) = H(\tilde{v}_i)$$
The entries of $\tilde{v}_i = (w_1,\dots,w_T)$ are substrings of $x$ of length $2^{k-i+1}$. Let
 $\Sigma \leq |\lambda|$ be the
positive constant such that
$$\mu(y) = \lambda(y)/\Sigma, \; \; y \in \{0,1\}^{2^{k-i+1}}$$
defines a probability distribution on $\{0,1\}^{2^{k-i+1}}$. Then,
\begin{eqnarray*}
H(\tilde{v}_i) & \leq & -\log_2\mu(w_1) -\log_2\mu(w_2) - \dots -\log_2\mu(w_T) \\
& \leq & |{S}_i|\log_2|\lambda| - \sum_{t=1}^T \log_2\lambda(w_t)
\end{eqnarray*}
Summing the preceding inequality over $i$ in the range $2 \leq i \leq k+1$, and using Property 3
together with the property (\ref{eq1000}) of $\lambda$, we obtain
(\ref{eq30}).\par

\begin{lemma} There is a sequence of positive numbers $\{\epsilon_k : k=1,2,\dots\}$ converging
to zero such that the following is true. For any $k = 1,2,\dots$ and any $G \in {\cal G}^*$
representing a binary string of length $2^k$, if we let $S_1, S_2, \dots, S_{k+1}$ be the strings
defined from $G$ in Section II,
\begin{equation}
|S_1| + |S_2| + \dots + |S_{k+1}| \leq \frac{2^{k+1}(2+\epsilon_k)}{k}
\label{eq44}
\end{equation}
\end{lemma}
\par

{\it Sketch of Proof.} Suppose  $G$ is any graph in $ {\cal G}^*$ representing a binary
string $x$ of length $2^{k}$. Let ${\cal S}(x)$ be the set
of all binary strings  which lie in the partitions of $x$ into substrings
of length $1, 2, 2^2, \dots, 2^k$. Define  the graph $G'$ to be the graph in which:
\begin{itemize}
\item The set of vertices of $G'$ is ${\cal S}(x)$. The set of terminal vertices of $G'$
is $\{0,1\}$.
\item For each nonterminal vertex $u$ of $G'$, there are two edges emanating from $u$, one
of which, labelled edge $0$, terminates at the left half of $u$, and the other
of which, labelled edge $1$, terminates at the right half of $u$.
\end{itemize}
The graph $G'$ is isomorphic to the graph termed by Liaw and Lin \cite{liaw} the {\it quasi-reduced} ordered
binary decision diagram corresponding to the ROBDD $G$. Let $V(G)$ be the set of vertices
of $G$ and let $V(G')$ be the set of vertices of $G'$. It is proved in the paper \cite{liaw}
that there exists a sequence of positive constants $\{\epsilon_k\}$ tending to zero
such that for any $k$ and any $G \in {\cal G}^*$ representing a binary string of
length $2^k$,
\begin{equation}
|V(G)| \leq |V(G')| \leq \frac{2^k(2+\epsilon_k)}{k}
\label{eq41}
\end{equation}
For the strings $S_1, S_2, \dots, S_{k+1}$ defined for this same $G$ as in Section II, it can
be shown (we omit the proof here) that
\begin{equation}
|S_1| + |S_2| + \dots + |S_{k+1}| \leq |V(G')| + |V(G)|
\label{eq42}
\end{equation}
Combining (\ref{eq41}) and (\ref{eq42}), we obtain (\ref{eq44}).
\par

Here is our main result.\par

\begin{theorem} Consider an arbitrary  binary $s$-state information source. For each binary
string $x$ of finite length, let $\mu(x)$ denote the probability assigned to $x$ by the given source.
Then, for $n = 2, 4, 8, 16, \dots$,
$$\max\{ x \in \{0,1\}^{n} \cap S({\rm dyadic}) : |\sigma(x)| + \log_2\mu(x)\} \leq
 \left ( \frac{n}{\log_2n}\right ) \left ( 16+4\log_2s + o(1)\right )$$
\end{theorem}
\par

{\it Proof.} Fix a $\lambda \in \Lambda$ such that
\begin{itemize}
\item $|\lambda| \leq s$.
\item $\mu(y) \leq \lambda(y)$ for every binary string $y$.
\end{itemize}
Fix $n \in \{1,2,4,8,\dots\}$ and $x \in \{0,1\}^n \cap S({\rm dyadic})$. Let $G \in {\cal G}^*$
be the graph $G = G_x$. Let $k = \log_2n$, and let $S_1, S_2, \dots, S_{k+1}$ be
the strings constructed from $G$ according to Section II.
Applying Lemmas 3 and 4 to (\ref{eq61}),
$$|\sigma(G)| \leq 4\left [ \frac{2^{k+1}(2+\epsilon_k)}{k}\right ] +
 \left [ \frac{2^{k+1}(2+\epsilon_k)}{k}\right ] \log_2s -\log_2\mu(x)$$
which gives us our result.  \newpage
\begin{center}
\setlength{\unitlength}{3947sp}%
\begingroup\makeatletter\ifx\SetFigFont\undefined%
\gdef\SetFigFont#1#2#3#4#5{%
  \reset@font\fontsize{#1}{#2pt}%
  \fontfamily{#3}\fontseries{#4}\fontshape{#5}%
  \selectfont}%
\fi\endgroup%
\begin{picture}(7981,8377)(1673,-8161)
\thinlines
\put(5101,-61){\circle{540}}
\put(3076,-1186){\circle{540}}
\put(7545,-1176){\circle{540}}
\put(6570,-2451){\circle{540}}
\put(8670,-2376){\circle{540}}
\put(4051,-2386){\circle{540}}
\put(5851,-3736){\circle{540}}
\put(7051,-3736){\circle{540}}
\put(9376,-3736){\circle{540}}
\put(8176,-3736){\circle{540}}
\put(5251,-4936){\circle{540}}
\put(4351,-6361){\circle{540}}
\put(1951,-7561){\circle{540}}
\put(7576,-7486){\circle{540}}
\put(2896,-4936){\circle{540}}
\put(2176,-2386){\circle{540}}
\put(4876,-211){\line(-2,-1){1590}}
\put(5401,-136){\line( 5,-2){1965.517}}
\put(2926,-1411){\line(-5,-6){645.492}}
\put(3301,-1336){\line( 5,-6){645.492}}
\put(7426,-1411){\line(-5,-6){713.115}}
\put(7801,-1261){\line( 1,-1){825}}
\put(2101,-2611){\line( 0,-1){4650}}
\put(2401,-2536){\line( 1,-2){1815}}
\put(3901,-2611){\line(-2,-5){822.414}}
\put(4201,-2611){\line( 1,-2){1005}}
\put(6451,-2686){\line(-2,-3){507.692}}
\put(6751,-2611){\line( 1,-3){300}}
\put(8626,-2611){\line(-2,-5){362.069}}
\put(8776,-2611){\line( 2,-3){542.308}}
\put(5626,-3886){\line(-1,-1){3450}}
\put(6001,-3961){\line( 1,-2){1620}}
\put(6901,-3961){\line(-1,-1){2287.500}}
\put(7126,-4036){\line( 1,-6){535.135}}
\put(8026,-3961){\line(-6,-1){4950}}
\put(8176,-3961){\line(-1,-6){537.162}}
\put(9226,-3961){\line(-4,-1){3741.177}}
\put(9376,-3961){\line(-1,-2){1650}}
\put(2776,-5236){\line(-1,-3){705}}
\put(3151,-5086){\line( 2,-1){4290}}
\put(5101,-5161){\line(-2,-3){657.692}}
\put(5476,-5086){\line( 1,-1){2100}}
\put(4201,-6586){\line(-2,-1){1920}}
\put(4576,-6511){\line( 3,-1){2767.500}}
\put(5026,-136){\makebox(0,0)[lb]{\smash{\SetFigFont{12}{14.4}{\rmdefault}{\mddefault}{\updefault}1}}}
\put(3001,-1261){\makebox(0,0)[lb]{\smash{\SetFigFont{12}{14.4}{\rmdefault}{\mddefault}{\updefault}2}}}
\put(7501,-1261){\makebox(0,0)[lb]{\smash{\SetFigFont{12}{14.4}{\rmdefault}{\mddefault}{\updefault}2}}}
\put(2101,-2461){\makebox(0,0)[lb]{\smash{\SetFigFont{12}{14.4}{\rmdefault}{\mddefault}{\updefault}3}}}
\put(3976,-2461){\makebox(0,0)[lb]{\smash{\SetFigFont{12}{14.4}{\rmdefault}{\mddefault}{\updefault}3}}}
\put(6526,-2536){\makebox(0,0)[lb]{\smash{\SetFigFont{12}{14.4}{\rmdefault}{\mddefault}{\updefault}3}}}
\put(8626,-2461){\makebox(0,0)[lb]{\smash{\SetFigFont{12}{14.4}{\rmdefault}{\mddefault}{\updefault}3}}}
\put(5776,-3811){\makebox(0,0)[lb]{\smash{\SetFigFont{12}{14.4}{\rmdefault}{\mddefault}{\updefault}4}}}
\put(8101,-3811){\makebox(0,0)[lb]{\smash{\SetFigFont{12}{14.4}{\rmdefault}{\mddefault}{\updefault}4}}}
\put(9301,-3811){\makebox(0,0)[lb]{\smash{\SetFigFont{12}{14.4}{\rmdefault}{\mddefault}{\updefault}4}}}
\put(6976,-3811){\makebox(0,0)[lb]{\smash{\SetFigFont{12}{14.4}{\rmdefault}{\mddefault}{\updefault}4}}}
\put(2851,-5011){\makebox(0,0)[lb]{\smash{\SetFigFont{12}{14.4}{\rmdefault}{\mddefault}{\updefault}5}}}
\put(5176,-5011){\makebox(0,0)[lb]{\smash{\SetFigFont{12}{14.4}{\rmdefault}{\mddefault}{\updefault}5}}}
\put(4276,-6436){\makebox(0,0)[lb]{\smash{\SetFigFont{12}{14.4}{\rmdefault}{\mddefault}{\updefault}6}}}
\put(1876,-7636){\makebox(0,0)[lb]{\smash{\SetFigFont{12}{14.4}{\rmdefault}{\mddefault}{\updefault}7}}}
\put(7501,-7561){\makebox(0,0)[lb]{\smash{\SetFigFont{12}{14.4}{\rmdefault}{\mddefault}{\updefault}7}}}
\put(1801,-8161){\makebox(0,0)[lb]{\smash{\SetFigFont{12}{14.4}{\rmdefault}{\mddefault}{\updefault}$T_G^0$}}}
\put(7426,-8086){\makebox(0,0)[lb]{\smash{\SetFigFont{12}{14.4}{\rmdefault}{\mddefault}{\updefault}$T_G^1$}}}
\end{picture}
\end{center}
\begin{center}
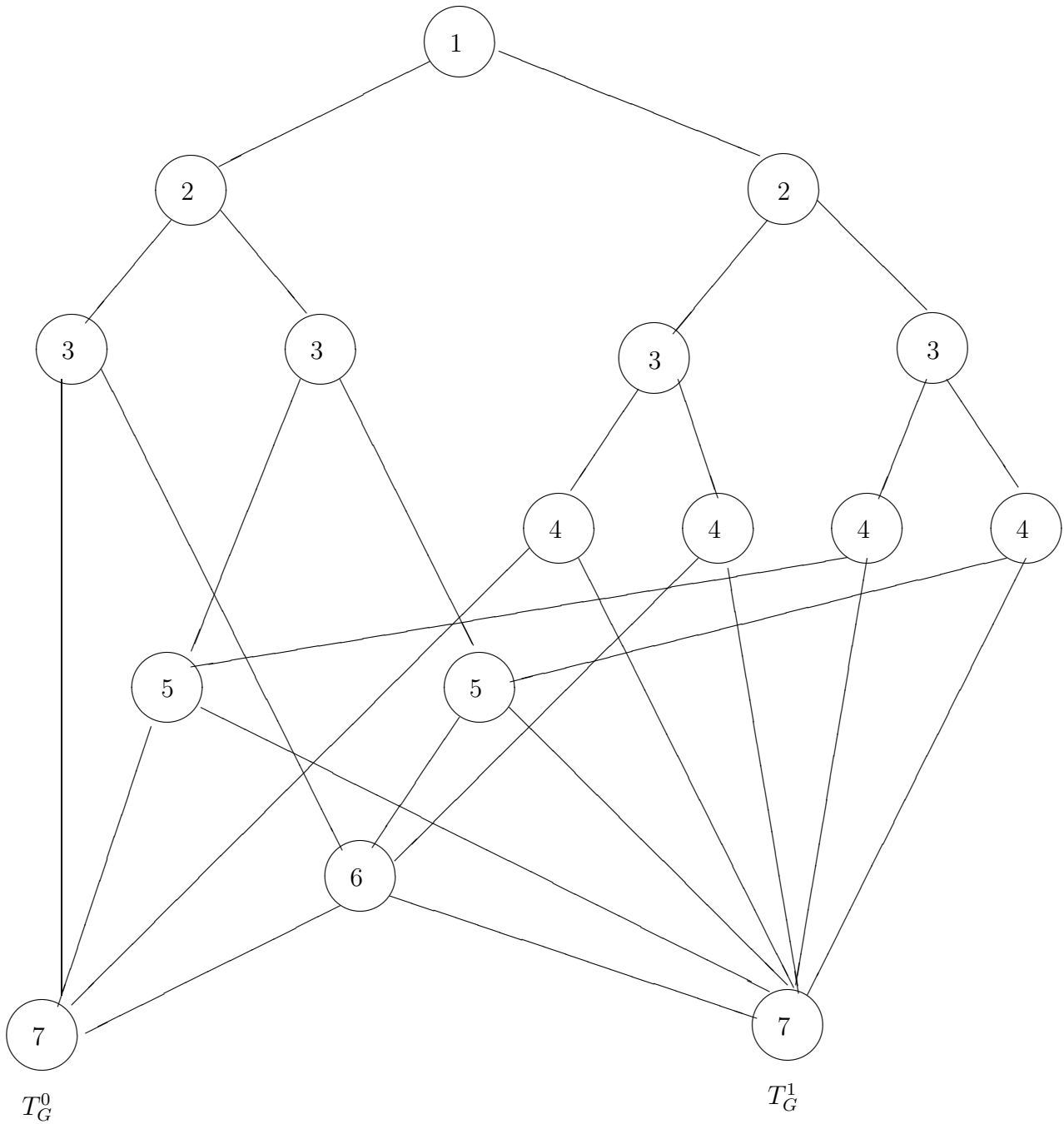
{\bf Figure 1:} A ROBDD $G$ from Bryant \cite{bryant1} (left edges labelled $0$, right edges labelled $1$)
\end{center}
\newpage
\begin{center}
\setlength{\unitlength}{3947sp}%
\begingroup\makeatletter\ifx\SetFigFont\undefined%
\gdef\SetFigFont#1#2#3#4#5{%
  \reset@font\fontsize{#1}{#2pt}%
  \fontfamily{#3}\fontseries{#4}\fontshape{#5}%
  \selectfont}%
\fi\endgroup%
\begin{picture}(8228,8235)(1426,-8011)
\thinlines
\put(5101,-61){\circle{540}}
\put(3076,-1186){\circle{540}}
\put(7545,-1176){\circle{540}}
\put(6570,-2451){\circle{540}}
\put(8670,-2376){\circle{540}}
\put(4051,-2386){\circle{540}}
\put(5851,-3736){\circle{540}}
\put(7051,-3736){\circle{540}}
\put(9376,-3736){\circle{540}}
\put(8176,-3736){\circle{540}}
\put(5251,-4936){\circle{540}}
\put(4351,-6361){\circle{540}}
\put(1951,-7561){\circle{540}}
\put(7576,-7486){\circle{540}}
\put(2896,-4936){\circle{540}}
\put(2176,-2386){\circle{540}}
\put(4876,-211){\line(-2,-1){1590}}
\put(5401,-136){\line( 5,-2){1965.517}}
\put(2926,-1411){\line(-5,-6){645.492}}
\put(3301,-1336){\line( 5,-6){645.492}}
\put(7426,-1411){\line(-5,-6){713.115}}
\put(7801,-1261){\line( 1,-1){825}}
\put(2101,-2611){\line( 0,-1){4650}}
\put(2401,-2536){\line( 1,-2){1815}}
\put(3901,-2611){\line(-2,-5){822.414}}
\put(4201,-2611){\line( 1,-2){1005}}
\put(6451,-2686){\line(-2,-3){507.692}}
\put(6751,-2611){\line( 1,-3){300}}
\put(8626,-2611){\line(-2,-5){362.069}}
\put(8776,-2611){\line( 2,-3){542.308}}
\put(5626,-3886){\line(-1,-1){3450}}
\put(6001,-3961){\line( 1,-2){1620}}
\put(6901,-3961){\line(-1,-1){2287.500}}
\put(7126,-4036){\line( 1,-6){535.135}}
\put(8026,-3961){\line(-6,-1){4950}}
\put(8176,-3961){\line(-1,-6){537.162}}
\put(9226,-3961){\line(-4,-1){3741.177}}
\put(9376,-3961){\line(-1,-2){1650}}
\put(2776,-5236){\line(-1,-3){705}}
\put(3151,-5086){\line( 2,-1){4290}}
\put(5101,-5161){\line(-2,-3){657.692}}
\put(5476,-5086){\line( 1,-1){2100}}
\put(4201,-6586){\line(-2,-1){1920}}
\put(4576,-6511){\line( 3,-1){2767.500}}
\put(5026,-136){\makebox(0,0)[lb]{\smash{\SetFigFont{12}{14.4}{\rmdefault}{\mddefault}{\updefault}1}}}
\put(3001,-1261){\makebox(0,0)[lb]{\smash{\SetFigFont{12}{14.4}{\rmdefault}{\mddefault}{\updefault}2}}}
\put(7501,-1261){\makebox(0,0)[lb]{\smash{\SetFigFont{12}{14.4}{\rmdefault}{\mddefault}{\updefault}2}}}
\put(2101,-2461){\makebox(0,0)[lb]{\smash{\SetFigFont{12}{14.4}{\rmdefault}{\mddefault}{\updefault}3}}}
\put(3976,-2461){\makebox(0,0)[lb]{\smash{\SetFigFont{12}{14.4}{\rmdefault}{\mddefault}{\updefault}3}}}
\put(6526,-2536){\makebox(0,0)[lb]{\smash{\SetFigFont{12}{14.4}{\rmdefault}{\mddefault}{\updefault}3}}}
\put(8626,-2461){\makebox(0,0)[lb]{\smash{\SetFigFont{12}{14.4}{\rmdefault}{\mddefault}{\updefault}3}}}
\put(5776,-3811){\makebox(0,0)[lb]{\smash{\SetFigFont{12}{14.4}{\rmdefault}{\mddefault}{\updefault}4}}}
\put(8101,-3811){\makebox(0,0)[lb]{\smash{\SetFigFont{12}{14.4}{\rmdefault}{\mddefault}{\updefault}4}}}
\put(9301,-3811){\makebox(0,0)[lb]{\smash{\SetFigFont{12}{14.4}{\rmdefault}{\mddefault}{\updefault}4}}}
\put(6976,-3811){\makebox(0,0)[lb]{\smash{\SetFigFont{12}{14.4}{\rmdefault}{\mddefault}{\updefault}4}}}
\put(2851,-5011){\makebox(0,0)[lb]{\smash{\SetFigFont{12}{14.4}{\rmdefault}{\mddefault}{\updefault}5}}}
\put(5176,-5011){\makebox(0,0)[lb]{\smash{\SetFigFont{12}{14.4}{\rmdefault}{\mddefault}{\updefault}5}}}
\put(4276,-6436){\makebox(0,0)[lb]{\smash{\SetFigFont{12}{14.4}{\rmdefault}{\mddefault}{\updefault}6}}}
\put(1876,-7636){\makebox(0,0)[lb]{\smash{\SetFigFont{12}{14.4}{\rmdefault}{\mddefault}{\updefault}7}}}
\put(7501,-7561){\makebox(0,0)[lb]{\smash{\SetFigFont{12}{14.4}{\rmdefault}{\mddefault}{\updefault}7}}}
\put(2926,-736){\makebox(0,0)[lb]{\smash{\SetFigFont{12}{14.4}{\rmdefault}{\mddefault}{\updefault}$A_2$}}}
\put(7801,-886){\makebox(0,0)[lb]{\smash{\SetFigFont{12}{14.4}{\rmdefault}{\mddefault}{\updefault}$A_3$}}}
\put(1801,-2011){\makebox(0,0)[lb]{\smash{\SetFigFont{12}{14.4}{\rmdefault}{\mddefault}{\updefault}$A_4$}}}
\put(4201,-1936){\makebox(0,0)[lb]{\smash{\SetFigFont{12}{14.4}{\rmdefault}{\mddefault}{\updefault}$A_5$}}}
\put(6376,-2011){\makebox(0,0)[lb]{\smash{\SetFigFont{12}{14.4}{\rmdefault}{\mddefault}{\updefault}$A_6$}}}
\put(8776,-1936){\makebox(0,0)[lb]{\smash{\SetFigFont{12}{14.4}{\rmdefault}{\mddefault}{\updefault}$A_7$}}}
\put(1426,-7261){\makebox(0,0)[lb]{\smash{\SetFigFont{12}{14.4}{\rmdefault}{\mddefault}{\updefault}$A_8$}}}
\put(3751,-6436){\makebox(0,0)[lb]{\smash{\SetFigFont{12}{14.4}{\rmdefault}{\mddefault}{\updefault}$A_9$}}}
\put(2551,-4561){\makebox(0,0)[lb]{\smash{\SetFigFont{12}{14.4}{\rmdefault}{\mddefault}{\updefault}$A_{10}$}}}
\put(5401,-4636){\makebox(0,0)[lb]{\smash{\SetFigFont{12}{14.4}{\rmdefault}{\mddefault}{\updefault}$A_{11}$}}}
\put(5551,-3361){\makebox(0,0)[lb]{\smash{\SetFigFont{12}{14.4}{\rmdefault}{\mddefault}{\updefault}$A_{12}$}}}
\put(6601,-3436){\makebox(0,0)[lb]{\smash{\SetFigFont{12}{14.4}{\rmdefault}{\mddefault}{\updefault}$A_{13}$}}}
\put(7876,-3361){\makebox(0,0)[lb]{\smash{\SetFigFont{12}{14.4}{\rmdefault}{\mddefault}{\updefault}$A_{14}$}}}
\put(9451,-3361){\makebox(0,0)[lb]{\smash{\SetFigFont{12}{14.4}{\rmdefault}{\mddefault}{\updefault}$A_{15}$}}}
\put(1651,-8011){\makebox(0,0)[lb]{\smash{\SetFigFont{12}{14.4}{\rmdefault}{\mddefault}{\updefault}$A_8 = T_G^0$}}}
\put(7276,-8011){\makebox(0,0)[lb]{\smash{\SetFigFont{12}{14.4}{\rmdefault}{\mddefault}{\updefault}$A_{16} = T_G^1$}}}
\put(5476, 89){\makebox(0,0)[lb]{\smash{\SetFigFont{12}{14.4}{\rmdefault}{\mddefault}{\updefault}$A_1 = V_G^r$}}}
\end{picture}
\end{center}
\begin{center}
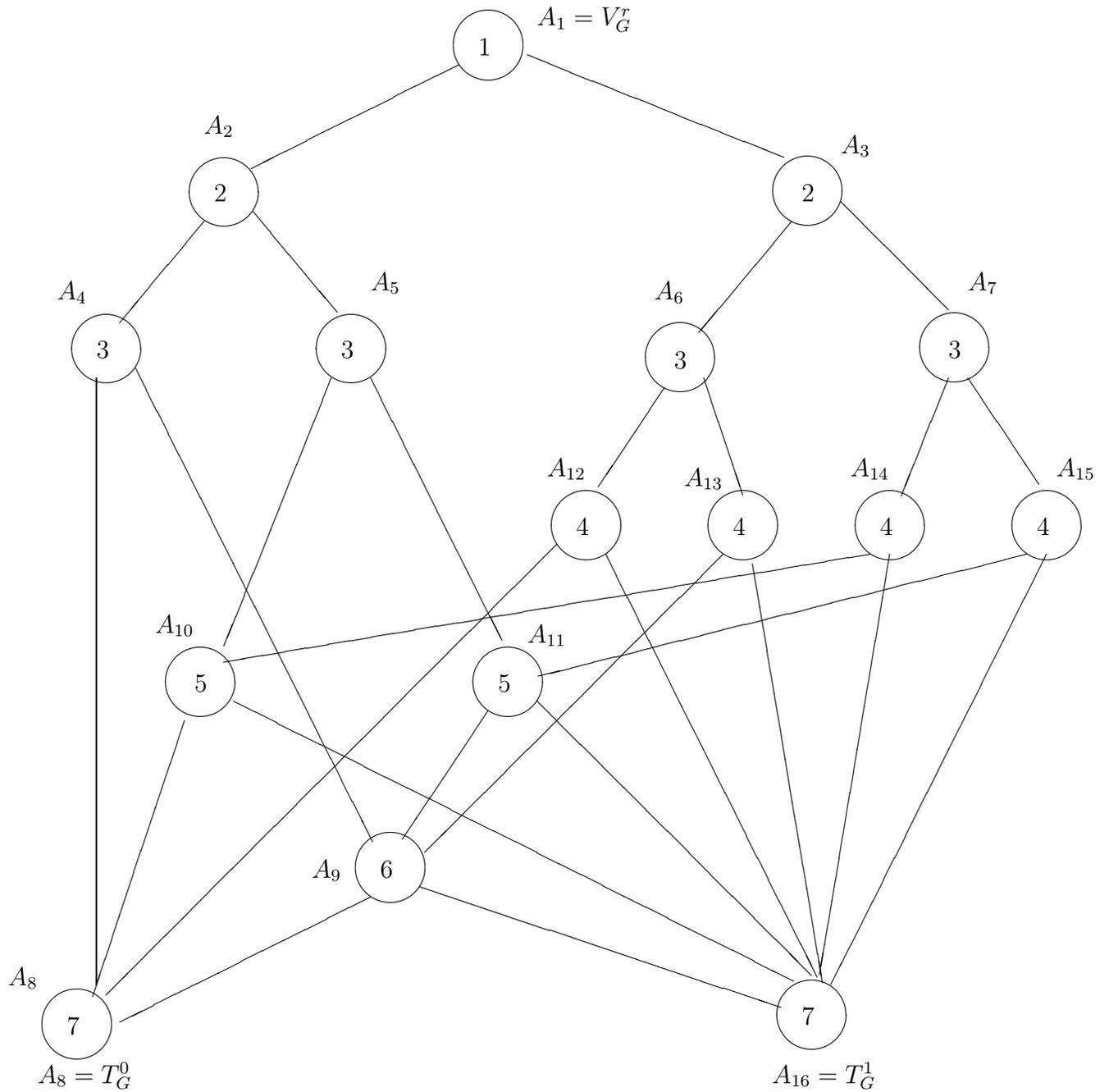
{\bf Figure 2:} Canonical ordering of vertices of ROBDD $G$ in Figure 1
\end{center}

\end{document}